\documentclass[amsmath,amsfonts,amssymb,superscriptaddress,preprint]{revtex4}
\usepackage{graphicx}
\usepackage{amsmath}
\usepackage{amsfonts}
\usepackage{amssymb}

\begin{document}

\title{ON BUILDING BETTER CONE JET ALGORITHMS\footnote{Contribution 
to the P5 Working Group on QCD and Strong Interactions at
Snowmass 2001}}
\author{S.D.\ Ellis}

\email[E-mail: ]{sdellis@u.washington.edu}

\affiliation{Department of Physics, University of Washington\\
Seattle, WA \ 98195 \ USA}

\author{J.\ Huston}

\email[E-mail: ]{huston@pa.msu.edu}

\affiliation{Department of Physics \&\ Astronomy,
Michigan State University\\
E. Lansing, MI 48824 \ USA}

\author{M.\ T\"{o}nnesmann}

\email[E-mail: ]{matthias@mppmu.mpg.de}

\affiliation{Max-Planck-Institut f\"{u}r Physik
(Werner-Heisenberg-Institut),\\
F\"{o}hringer Ring 6, 80805 M\"{u}nchen, Germany}

\date{November 28, 2001}
\begin{abstract}
We discuss recent progress in understanding the issues essential to the
development of better cone jet algorithms.
\end{abstract}

\preprint{UW/PT-01-25}
\preprint{MSUHEP-11101}
\preprint{MPI-PhE/2001-16}

\maketitle



%


%

An important facet of preparations\cite{workshops} for Run II\ at the
Tevatron, and for future data taking at the LHC, has been the study of ways in
which to improve jet algorithms. \ These algorithms are employed to map final
states, both in QCD perturbation theory and in the data, onto jets. \ The
motivating idea is that these jets are the surrogates for the underlying
energetic partons. \ In principle, we can connect the observed final states,
in all of their complexity, with the perturbative final states, which are
easier to interpret and to analyze theoretically. \ Of necessity these jet
algorithms should be robust under the impact of both higher order perturbative
and non-perturbative physics and the effects introduced by the detectors
themselves. \ The quantitative goal is a precision of order 1\% in the mapping
between theory and experiment. \ In this note we will provide a brief summary
of recent progress towards this goal. \ A more complete discussion of our
results will be provided elsewhere\cite{longpaper}. \ Here we will focus on
cone jet algorithms, which have formed the basis of jet studies at hadron colliders.

As a starting point we take the Snowmass Algorithm\cite{Snowmass}, which was
defined by a collaboration of theorists and experimentalists and formed the
basis of the jet algorithms used by the CDF and D\O \ collaborations during
Run I at the Tevatron. \ Clearly jets are to be composed of either hadrons or
partons that\ are, in some sense, nearby each other. \ The cone jet defines
nearness in an intuitive geometric fashion: jets are composed of hadrons or
partons whose 3-momenta lie within a cone defined by a circle in $\left(
\eta,\phi\right)  $. \ These are essentially the usual angular variables,
where $\eta=\ln\left(  \cot\,\theta/2\right)  $ is the pseudorapidity and
$\phi$ is the azimuthal angle. This idea of being nearby in angle can be
contrasted with an algorithm based on being nearby in transverse momentum as
illustrated by the so-called $k_{T}$ Algorithm\cite{ktcluster} that has been
widely used at $e^{+}e^{-}$ and $ep$ colliders. \ We also expect the jets to
be aligned with the most energetic particles in the final state. \ This
expectation is realized in the Snowmass Algorithm by defining an acceptable
jet in terms of a ``stable'' cone such that the geometric center of the cone
is identical to the $E_{T\text{ }}$ weighted centroid. \ Thus, if we think of
a sum over final state partons or hadrons defined by an index $k$ and in the
direction $\left(  \eta_{k},\phi_{k}\right)  $, a jet ($J$) of cone radius $R$
is defined by the following set of equations%

\begin{align}
k  &  \in J:\left(  \phi_{k}-\phi_{J}\right)  ^{2}+\left(  \eta_{k}-\eta
_{J}\right)  ^{2}\leq R^{2},\nonumber\\
\phi_{J}  &  =\sum_{k\in J}\frac{E_{T,k}\phi_{k}}{E_{T,J}},\quad\eta_{J}%
=\sum_{k\in J}\frac{E_{T,k}\eta_{k}}{E_{T,J}},\label{cone}\\
E_{T,J}  &  =\sum_{k\in J}E_{T,k}.\nonumber
\end{align}
In these expressions $E_{T}$ is the transverse energy ($\left|
\overrightarrow{p}_{T}\right|  $ for a massless 4-vector). \ It is important
to recognize that jet algorithms involve two distinct steps. \ The first step
is to identify the ``members'' of the jet, \textit{i.e}., the calorimeter
towers or the partons that make-up the stable cone that becomes the jet. \ The
second step involves constructing the kinematic properties that will
characterize the jet, \textit{i.e.}, determine into which bin the jet will be
placed. \ In the original Snowmass Algorithm the $E_{T}$ weighted variables
defined in Eq. \ref{cone} are used both to identify and bin the jet.

In a theoretical calculation one integrates over the phase space corresponding
to parton configurations that satisfy the stability conditions. \ In the
experimental case one searches for sets of final state particles (and
calorimeter towers) in each event that satisfy the constraint. \ In
practice\cite{workshops} the experimental implementation of the cone algorithm
has involved the use of various short cuts to minimize the search time. \ In
particular, Run I algorithms made use of seeds. \ Thus one looks for stable
cones only in the neighborhood of calorimeter cells, the seed cells, where the
deposited energy exceeds a predefined limit. \ Starting with such a seed cell,
one makes a list of the particles (towers) within a distance $R$ of the seed
and calculates the centroid for the particles in the list (calculated as in
Eq. \ref{cone}). \ If the calculated centroid is consistent with the initial
cone center, a stable cone has been identified. \ If not, the calculated
centroid is used as the center of a new cone with a new list of particles
inside and the calculation of the centroid is repeated. \ This process is
iterated, with the cone center migrating with each repetition, until a stable
cone is identified or until the cone centroid has migrated out of the fiducial
volume of the detector. \ When all of the stable cones in an event have been
identified, there will typically be some overlap between cones. \ This
situation must be addressed by a splitting/merging routine in the jet
algorithm. \ This feature was not foreseen in the original Snowmass Algorithm.
\ Normally this involves the definition of a parameter $f_{merge}$, typically
with values in the range $0.5\ \leq f_{merge}\leq0.75,$ such that, if the
overlap transverse energy fraction (the transverse energy in the overlap
region divided by the smaller of the total energies in the two overlapping
cones) is greater than $f_{merge}$, the two cones are merged to make a single
jet. If this constraint is not met, the calorimeter towers/hadrons in the
overlap region are individually assigned to the cone whose center is closer.
\ This situation yields 2 final jets. 

The essential challenge in the use of jet algorithms is to understand the
differences between the experimentally applied algorithms and the
theoretically applied ones and hence understand the uncertainties. \ This is
the primary concern of this paper. \ It has been known for some time that the
use of seeds in the experimental algorithms means that certain configurations
kept by the theoretical algorithm are likely to be missed by the experimental
one\cite{SDERsep}. \ At higher orders in perturbation theory the seed
definition also introduces an undesirable (logarithmic) dependence on the seed
$E_{T}$ cut (the minimum $E_{T}$ required to be treated as a seed
cell)\cite{Seymour}. \ Various alternative algorithms are described in the Run
II Workshop proceedings\cite{workshops} for addressing this issue, including
the Midpoint Algorithm and the Seedless Algorithm. \ In the last year it has
also been recognized that other final state configurations are likely to be
missed in the data, compared to the theoretical result. \ In this paper we
will explain these new developments and present possible solutions. \ To see
that there is a problem, we apply representative jet algorithms to data sets
that were generated with the HERWIG Monte Carlo\cite{Herwig} and then run
through a CDF detector simulation. \ As a reference we include in our analysis
the JetClu Algorithm\cite{JetClu}, which is the algorithm used by CDF in Run
I. \ It employs both seeds and a property called ``ratcheting''. \ This latter
term labels the fact that the Run I CDF algorithm (unlike the corresponding
D\O \ algorithm) was defined so that calorimeter towers initially found in a
cone around a seed continue to be associated with that cone, even as the
center of the cone migrates due to the iteration of the cone algorithm. \ Thus
the final ``footprint'' of the cone is not necessarily a circle in $\left(
\eta,\phi\right)  $ (even before the effects of splitting/merging). \ Since
the cone is ``tied'' to the initial seed towers, this feature makes it
unlikely that cones will migrate very far before becoming stable. We describe
results from JetClu both with and without this ratcheting feature. \ The
second cone algorithm studied is the Midpoint Algorithm that, like the JetClu
Algorithm, starts with seeds to find stable cones (but without ratcheting).
The Midpoint Algorithm then adds a cone at the midpoint in $\left(  \eta
,\phi\right)  $ between all identified pairs of stable cones separated by less
than $2R$ and iterates this cone to test for stability. \ This step is meant
to ensure that no stable ``mid-cones'' are missed, compared to the theoretical
result, due to the use of seeds. \ Following the recommendation of the Run
II\ Workshop, we actually use 4-vector kinematics for the Midpoint Algorithm
and place the cone at the midpoint in $\left(  y,\phi\right)  $, where $y$ is
the true rapidity. \ The third cone algorithm is the Seedless Algorithm that
places an initial trial cone at every point on a regular lattice in $\left(
y,\phi\right)  $, which is approximately as fine-grained as the detector. \ It
is not so much that this algorithm lacks seeds, but rather that the algorithm
puts seed cones ``everywhere''. \ The Seedless Algorithm can be streamlined by
imposing the constraint that a given trial cone is removed from the analysis
if the center of the cone migrates outside of its original lattice cell during
the iteration process. \ The streamlined version still samples every lattice
cell for stable cone locations, but is less computationally intensive. \ Our
experience with the streamlined version of this algorithm suggests that there
can be problems finding stable cones with centers located very close to cell
boundaries. \ This technical difficulty is easily addressed by enlarging the
distance that a trial cone must migrate before being discarded. \ For example,
if this distance is 60\% of the lattice cell width instead of the default
value of 50\%, the problem essentially disappears with only a tiny impact on
the required time for analysis. \ In the JetClu Algorithm the value
$f_{merge}=0.75$ was used (as in the Run I\ analyses), while for the other two
cone algorithms the value $f_{merge}=0.5$ was used as suggested in the
Workshop Proceedings\cite{workshops}. \ \ Finally, for completeness, we
include in our analysis a sample\ $k_{T}$ Algorithm. \ 

\begin{figure}
[ptb]
\begin{center}
\includegraphics[
height=6.7006in,
width=6.7006in
]%
{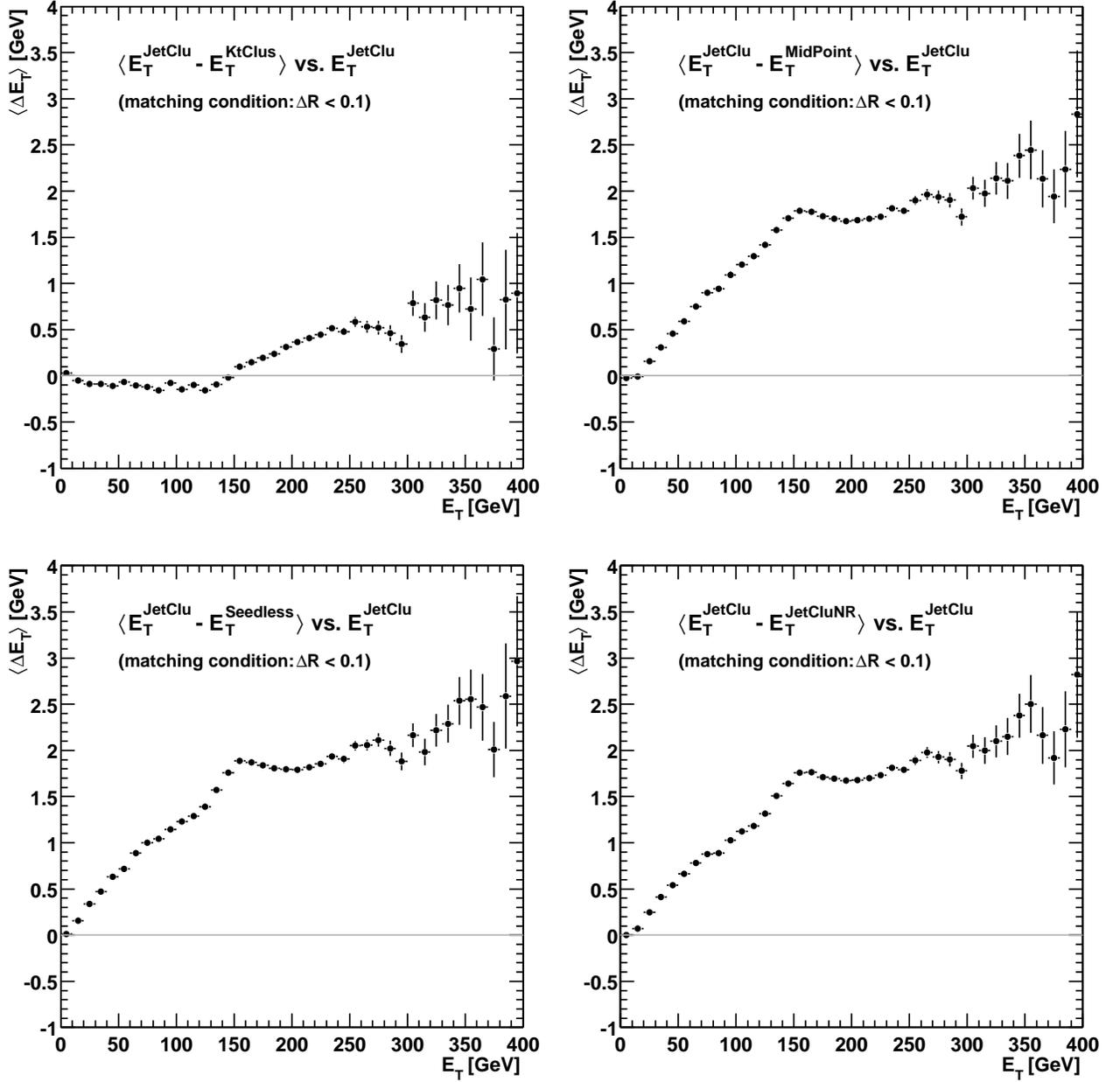}%
\caption{Difference of $E_{T}$ for matched jets found with various jet
algorithms and compared to the JetClu CDF Run I\ algorithm. \ The events
studied were generated with HERWIG 6.1 and run through the CDF detector
simulation.}%
\label{deltaET}%
\end{center}
\end{figure}

Starting with a sample of 250,000 events, which were generated with HERWIG 6.1
and run through a CDF\ detector simulation and which were required to have at
least 1 initial parton with $E_{T}>200$ GeV, we applied the various algorithms
to find jets with $R=0.7$ in the central region ($\left|  \eta\right|  <1$).
\ We then identified the corresponding jets from each algorithm by finding jet
centers differing by $\Delta R<0.1$. \ The plots in Fig. \ref{deltaET}
indicate the average difference in $E_{T}$ for these jets as a function of the
jet $E_{T}$. \ (We believe that some features of the indicated structure, in
particular the ``knees'' near $E_{T}=150$ GeV, are artifacts of the event
selection process.)%
From these results we can draw several conclusions. \ First, the $k_{T}$
Algorithm identifies jets with $E_{T}$ values similar to those found by
JetClu, finding slightly more energetic jets at small $E_{T}$ and somewhat
less energetic jets at large $E_{T}$. \ We will not discuss this algorithm
further here except to note that D\O \ has applied it in a study of Run I
data\cite{D0} and in that analysis the $k_{T}$ Algorithm jets seems to exhibit
slightly \emph{larger} $E_{T}$ than expected from NLO perturbation theory. The
cone algorithms, including the JetClu Algorithm without ratcheting, which is
labeled JetCluNR, identify jets with approximately 0.5\% to 1 \%
\emph{smaller} $E_{T}$ values than those identified by the JetClu Algorithm
(with ratcheting), with a corresponding approximately 5\% smaller jet cross
section at a given $E_{T}$ value. \ We believe that this systematic shortfall
can be understood as resulting from the smearing effects of perturbative
showering and non-perturbative hadronization. \ 

To provide insight into the issues raised by Fig. \ref{deltaET} we now
discuss\ a simple, but informative analytic picture. \ It will serve to
illustrate the impact of showering and hadronization on the operation of jet
algorithms. \ We consider the scalar function $F\left(  \overrightarrow
{r}\right)  $ defined as a function of the 2-dimensional variable
$\overrightarrow{r}=\left(  \eta,\phi\right)  $ by the integral over the
transverse energy distribution of either the partons or the
hadrons/calorimeter towers in the final state with the indicated weight function,%

\begin{align}
F\left(  \overrightarrow{r}\right)   &  =\frac{1}{2}\int d^{2}\rho
\times\left(  R^{2}-\left(  \overrightarrow{\rho}-\overrightarrow{r}\right)
^{2}\right)  \times\,\Theta\left(  R^{2}-\left(  \overrightarrow{\rho
}-\overrightarrow{r}\right)  ^{2}\right)  \times E_{T}\left(  \overrightarrow
{\rho}\right) \label{scalar}\\
&  =\frac{1}{2}\sum_{i}E_{T,i}\times\left(  R^{2}-\left(  \overrightarrow
{\rho_{i}}-\overrightarrow{r}\right)  ^{2}\right)  \times\,\Theta\left(
R^{2}-\left(  \overrightarrow{\rho_{i}}-\overrightarrow{r}\right)
^{2}\right)  .\nonumber
\end{align}
\newline The second expression arises from replacing the continuous energy
distribution with a discrete set, $i=1$ to $N,$ of delta functions,
representing the contributions of either a configuration of partons or a set
of calorimeter towers (and hadrons). \ Each parton direction or the location
of the center of each calorimeter tower is defined in $\eta$, $\phi$ by
$\rho_{i}=\left(  \eta_{i},\phi_{i}\right)  $, while the parton/calorimeter
cell has a transverse energy (or $E_{T}$) content given by $E_{T,i}$. \ This
function is clearly related to the energy in a cone of size $R$ containing the
towers whose centers lie within a circle of radius $R$ around the point
$\overrightarrow{r}$. \ More importantly it carries information about the
locations of ``stable'' cones. \ The points of equality between the $E_{T}$
weighted centroid and the geometric center of the cone correspond precisely to
the maxima of $F$. \ The gradient of this function has the form (note that the
delta function arising from the derivative of the theta function cannot
contribute as it is multiplied by a factor equal to its argument)%

\begin{equation}
\overrightarrow{\nabla}F\left(  \overrightarrow{r}\right)  =\sum_{i}%
E_{T,i}\times\left(  \overrightarrow{\rho_{i}}-\overrightarrow{r}\right)
\times\Theta\left(  R^{2}-\left(  \overrightarrow{\rho_{i}}-\overrightarrow
{r}\right)  ^{2}\right)  . \label{gradient}%
\end{equation}
This expression vanishes at points where the weighted centroid coincides with
the geometric center, \textit{i.e.}, at points of stability (and at minima of
$F$, points of extreme instability). \ The corresponding expression for the
energy in the cone centered at $\overrightarrow{r}$ is
\begin{equation}
E_{C}\left(  \overrightarrow{r}\right)  =\sum_{i}E_{T,i}\times\Theta\left(
R^{2}-\left(  \overrightarrow{\rho_{i}}-\overrightarrow{r}\right)
^{2}\right)  . \label{energy}%
\end{equation}%

\begin{figure}
[ptb]
\begin{center}
\includegraphics[
natheight=5.955100in,
natwidth=7.726200in,
height=3.8553in,
width=6.7974in
]%
{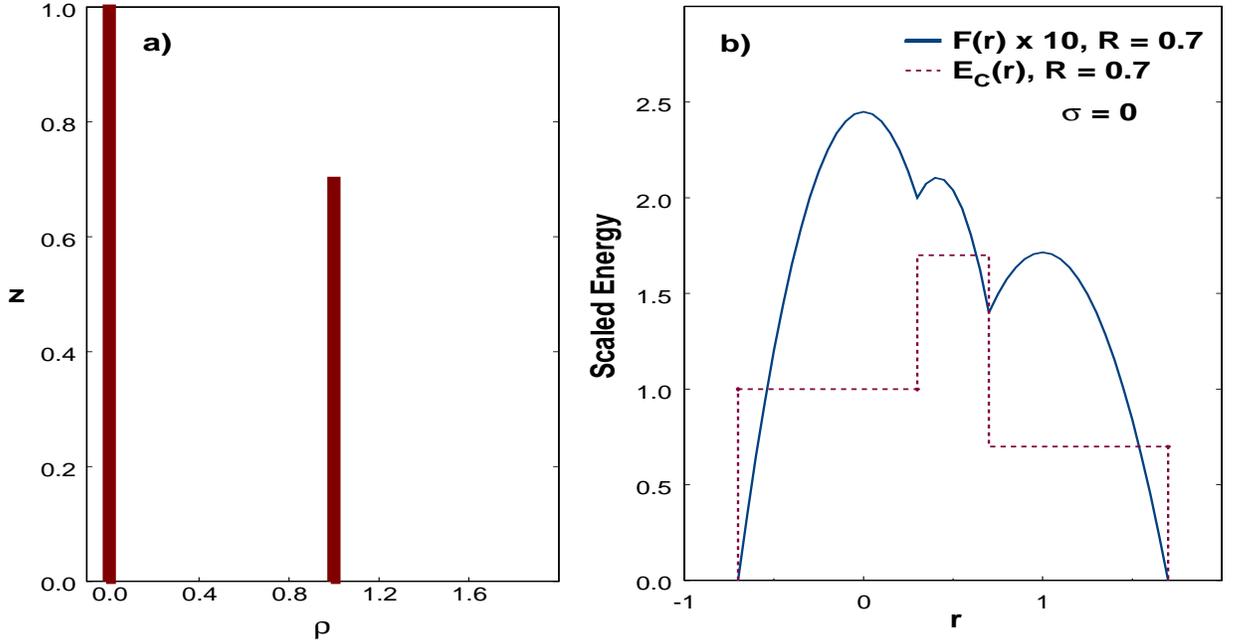}%
\caption{2-Parton distribution: a) transverse energy distribution; b)
distributions \textit{F(r)} and \textit{E}$_{C}$\textit{(r)} in the
perturbative limit of no smearing.}%
\label{edist2}%
\end{center}
\end{figure}
\ 

To more easily develop our understanding of these equations consider a
simplified scenario (containing all of the interesting effects) involving 2
partons separated in just one angular dimension $\overrightarrow{\rho
}\rightarrow\rho$ ($\overrightarrow{r}\rightarrow r$) with $\rho_{2}-\rho
_{1}=d$. \ It is sufficient to specify the energies of the 2 partons simply by
their ratio, $z=E_{2}/E_{1}\leq1$. \ Now we can study what sorts of 2 parton
configurations yield stable cones in this 2-D phase space specified by $0\leq
z\leq1$, $0\leq d\leq2R$ (beyond $2R$ the 2 partons are surely in different
cones). \ As a specific example consider the case $\rho_{1}=0$, $\rho
_{2}=d=1.0$ and $z=0.7$ with $R=0.7$ (the typical experimental value). \ The
underlying energy distribution is illustrated in Fig.\ \ref{edist2}a,
representing a delta function at $\rho=0$ (with scaled weight 1) and another
at $\rho=1.0$ (with scaled weight 0.7). \ This simple distribution leads to
the functions $F(r)$ and $E_{C}(r)$ indicated in Fig.\ \ref{edist2}b. \ In
going from the true energy distribution to the distribution $E_{C}(r)$ the
energy is effectively smeared over a range given by $R$. \ In $F(r)$ the
distribution is further shaped by the quadratic factor $R^{2}-\left(  \rho
_{i}-r\right)  ^{2}$. \ We see that $F(r)$ exhibits 3 local maxima
corresponding to the expected stable cones around the two original delta
functions ($r_{1}=0,r_{2}=1$), plus a third stable cone in the middle
($r_{3}=zd/\left(  1+z\right)  =0.41$ in the current case). \ This middle cone
includes the contributions from both partons as indicated by the magnitude of
the middle peak in the function $E_{C}(r)$. \ Note further that the middle
cone is found at a location where there is initially no energy in
Fig.\ \ref{edist2}a, and thus no seeds. \ One naively expects that such a
configuration is not identified as a stable cone by the experimental
implementations of the cone algorithm that use seeds simply because they do
not look for it. \ Note also that, since both partons are entirely within the
center cone, the overlap fractions are unity and the usual merging/splitting
routine will lead to a single jet containing all of the initial energy
($1+z$). \ This is precisely how this configuration was treated in the
NLO\ perturbative analysis of the Snowmass Algorithm\cite{EKS} (\textit{i.e}.,
only the leading jet, the middle cone, was kept). \ 

\begin{figure}
[ptb]
\begin{center}
\includegraphics[
natheight=5.955100in,
natwidth=7.726200in,
height=3.5561in,
width=6.4325in
]%
{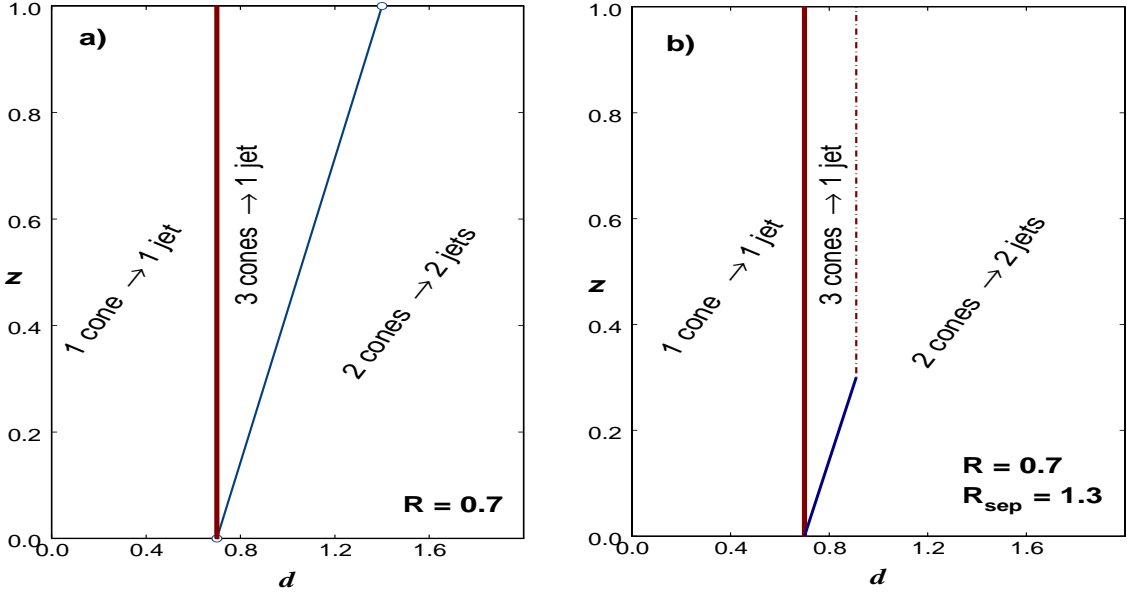}%
\caption{Perturbation Theory Structure: a) $R_{sep}=2$; b) $R_{sep}=1.3$.}%
\label{perthy4}%
\end{center}
\end{figure}

Similar reasoning leads to Fig.\ \ref{perthy4}a, which indicates the various 2
parton configurations found by the perturbative cone algorithm. \ For $d<R$
one finds a single stable cone and a single jet containing both partons. \ For
$R<d<(1+z)R$ one finds 3 stable cones that merge to 1 jet, again with all of
the energy. \ For $d>(1+z)R$ \ we find 2 stable cones and 2 jets, each
containing one parton, of scaled energies $1$ and $z$. \ Thus, except in the
far right region of the graph, the 2 partons are always merged to form a
single jet.%
We expect that the impact of seeds in experimental algorithms can be (crudely)
simulated in the NLO calculations\cite{SDERsep} by including a parameter
$R_{sep}$ such that stable cones containing 2 partons are not allowed for
partons separated by $d>R_{sep}\times R$.\ \ As a result cones are no longer
merged in this kinematic region. \ In the present language this situation is
illustrated in Fig.\ \ref{perthy4}b corresponding to $R_{sep}=1.3$, $R\times
R_{sep}=0.91$. \ This specific value for $R_{sep}$ was chosen\cite{SDERsep} to
yield reasonable agreement with the Run I data. \ The conversion of much the 3
cones $\rightarrow$ 1 jet region to 2 cones $\rightarrow$ 2 jets has the
impact of lowering the average $E_{T}$ of the leading jet and hence the jet
cross section at a fixed $E_{T,J}$. \ Parton configurations that naively
produced jets with energy characterized by $1+z$ now correspond to jets of
maximum energy 1. \ This is just the expected impact of a jet algorithm with
seeds. \ Note that with this value of $R_{sep}$ the specific parton
configuration in Fig. \ref{edist2}a will yield 2 jets (and not 1 merged jet)
in the theoretical calculation.\ \ As mentioned earlier this issue is to be
addressed by the Midpoint and Seedless Algorithms in Run II. \ However, as
indicated in Fig. \ref{deltaET}, neither of these two algorithms reproduces
the results of JetClu. \ Further, they both identify jets that are similar to
JetClu \emph{without} ratcheting. \ Thus we expect that there is more to this story.

\begin{figure}
[ptb]
\begin{center}
\includegraphics[
natheight=5.955100in,
natwidth=7.726200in,
height=3.4376in,
width=6.7758in
]%
{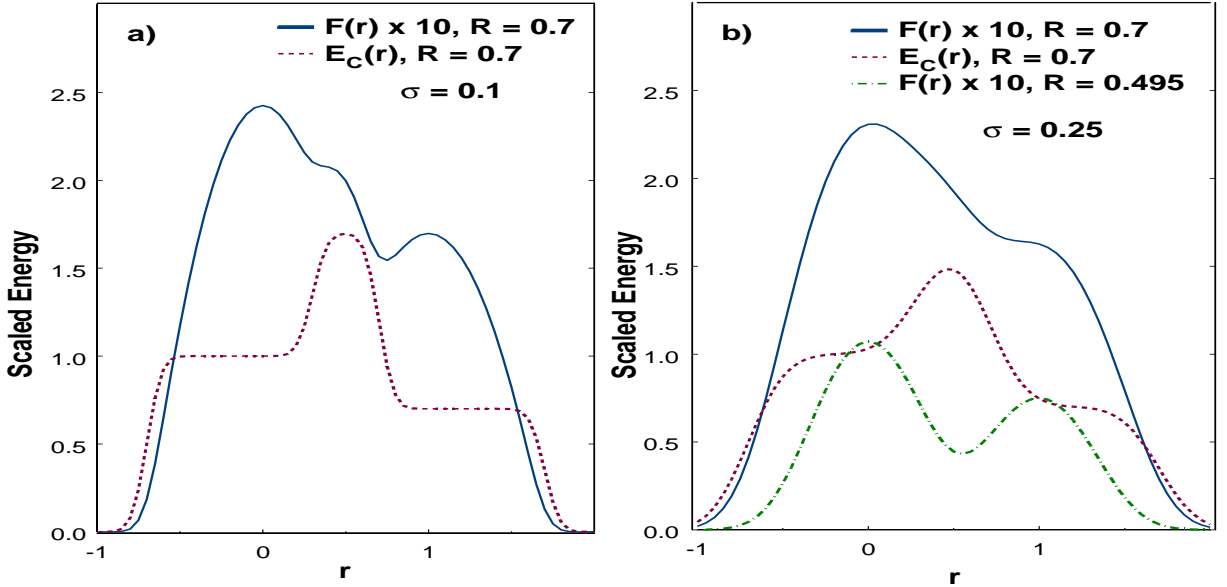}%
\caption{The distributions \textit{F(r)} and \textit{E}$_{C}$\textit{(r) }for
smearing width a) 0.1; b) 0.25.}%
\label{smearpt5}%
\end{center}
\end{figure}

As suggested earlier, a major difference between the perturbative level, with
a small number of partons, and the experimental level of multiple hadrons is
the smearing that results from perturbative showering and nonperturbative
hadronization. \ For the present discussion the primary impact is that the
starting energy distribution will be smeared out in the variable $r$. \ We can
simulate this effect in our simple model using gaussian smearing,
\textit{i.e}., we replace the delta functions in Eq. \ref{scalar} with
gaussians of width $\sigma$. \ (Since this corresponds to smearing in an
angular variable, we would expect $\sigma$ to be a decreasing function of
$E_{T}$, \textit{i.e}., more energetic jets are narrower. \ We also note that
this naive picture does not include the expected color coherence in the
products of the showering/hadronization process.) \ The first impact of this
smearing is that some of the energy initially associated with the partons now
lies outside of the cones centered on the partons. \ This effect, typically
referred to as ``splashout'' in the literature, is (exponentially) small in
this model for $\sigma<R$. \ Here we will focus on less well known but
phenomenologically more relevant impacts of splashout.\ The distributions
corresponding to Fig.\ \ref{edist2}b, but now with $\sigma=0.10$ (instead of
$\sigma=0$), are exhibited in Fig.\ \ref{smearpt5}a.%
\ With the initial energy distribution smeared by $\sigma$, the distribution
$F(r)$ is now even more smeared and, in fact, we see that the middle stable
cone (the maximum in the middle of Fig.\ \ref{edist2}b) has been washed out by
the increased smearing. \ Thus the cone algorithm applied to data (where such
smearing is present) may not find the middle cone that is present in
perturbation theory, not only due to the use of seeds but also due to this new
variety of splashout correction, which renders this cone unstable. \ Since, as
a result of this splashout correction, the middle cone\ is not stable, this
problem is \emph{not} addressed by either the Midpoint Algorithm or the
Seedless Algorithm. \ Both algorithms may look in the correct place, but they
look for stable cones. \ This point is presumably part of the explanation for
why both of these algorithms disagree with the JetClu results in Fig.
\ref{deltaET}.

\begin{figure}
[ptb]
\begin{center}
\includegraphics[
height=3.4298in,
width=4.8983in
]%
{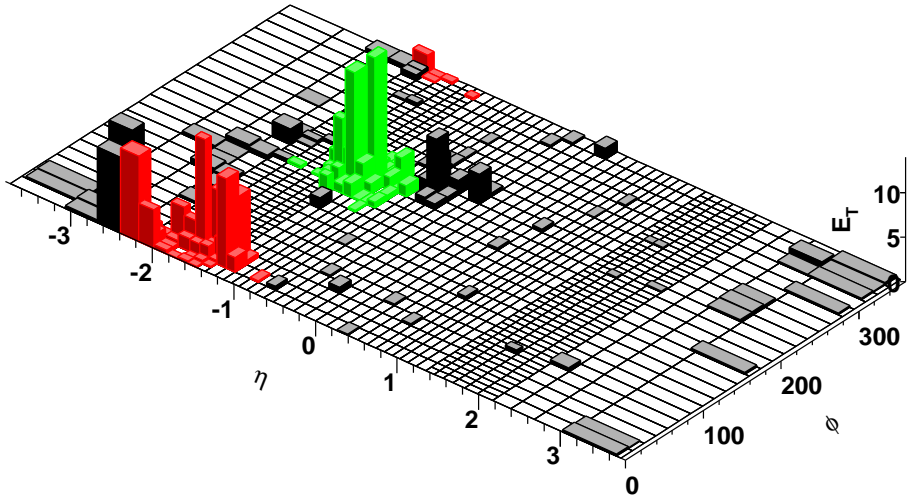}%
\caption{Result of applying the Midpoint Algorithm to a specific Monte Carlo
event in the CDF detector.}%
\label{lego}%
\end{center}
\end{figure}

Our studies also suggest a further impact of the smearing of
showering/hadronization that was previously unappreciated. \ This new effect
is illustrated in Fig.\ \ref{smearpt5}b, which shows $F(r)$, still for $z=0.7$
and $d=1.0,$ but now for $\sigma=0.25$. \ With the increased smearing the
second stable cone, corresponding to the second parton, has now also been
washed out, \textit{i.e}., the right hand local maximum\ has also disappeared.
\ This situation is exhibited in the case of ``data'' by the lego plot in Fig.
\ref{lego} indicating the jets found by the Midpoint Algorithm in a specific
Monte Carlo event. \ The Midpoint Algorithm does not identify the energetic
towers (shaded in black) to the right of the energetic central jet as either
part of that jet or as a separate jet, \textit{i.e}., these obviously relevant
towers are not found to be in a stable cone. \ The iteration of any cone
containing these towers invariably migrates to the nearby higher $E_{T}$
towers.%
\ 

\begin{figure}
[ptb]
\begin{center}
\includegraphics[
height=3.1064in,
width=3.1064in
]%
{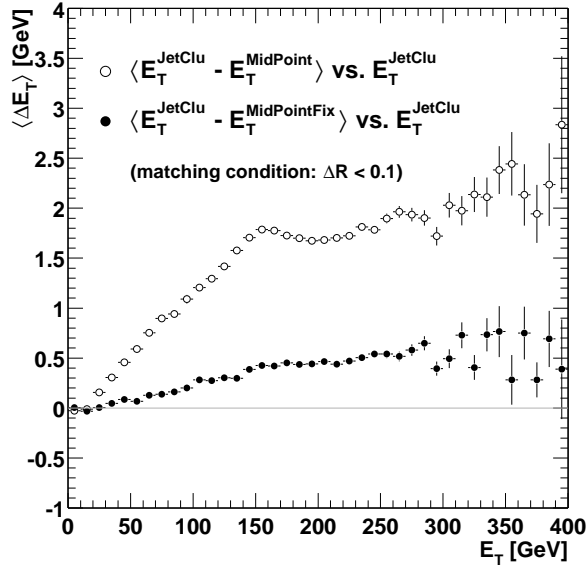}%
\caption{The difference in the $E_{T}$ of identified central jets for the
JetClu and Midpoint Algorithms, both with and without the ``fix'' discussed in
the text. \ The events studied were generated with HERWIG 6.1 and run through
the CDF detector simulation.}%
\label{deltaETfix}%
\end{center}
\end{figure}

In summary, we have found that the impact of smearing and splashout is
expected to be much more important than simply the leaking of energy out of
the cone. \ Certain stable cone configurations, present at the perturbative
level, can disappear from the analysis of real data due to the effects of
showering and hadronization. \ This situation leads to corrections to the
final jet yields that are relevant to our goal of 1\% precision in the mapping
between perturbation theory and experiment. \ Compared to the perturbative
analysis of the 2-parton configuration, both the middle stable cone and the
stable cone centered on the lower energy parton can be washed out by smearing.
\ Further, this situation is not addressed by either the Midpoint Algorithm or
the Seedless Algorithm. \ One possibility for addressing the missing middle
cone would be to eliminate the stability requirement for the added midpoint
cone in the Midpoint Algorithm. \ However, if there is enough smearing to
eliminate also the second (lower energy) cone, even this scenario will not
help, as we do not find two cones to put a third cone between. \ There is, in
fact, a rather simple ``fix'' that can be applied to the Midpoint Algorithm to
address this latter form of the splashout correction. \ We can simply use 2
values for the cone radius $R$, one during the search for the stable cones and
the second during the calculation of the jet properties. \ As a simple
example, the 3rd curve in Fig.\ \ref{smearpt5}b corresponds to using
$R/\sqrt{2}=0.495$ during the stable cone discovery phase and $R=0.7$ in the
jet construction phase. \ Thus the $R/\sqrt{2}$ value is used only during
iteration; the cone size is set to $R$ right after the stable cones have been
identified and the larger cone size is employed during the splitting/merging
phase. \ By comparing Figs.\ \ref{smearpt5}b and\ \ref{edist2}b we see that
the two outer stable cones in the perturbative case are in essentially the
same locations as in the smeared case using the smaller cone during discovery.
\ The improved agreement between the JetClu results and those of the Midpoint
Algorithm with the last ``fix'' (using the smaller $R$\ value during
discovering but still requiring cones to be stable) are indicated in Fig.
\ref{deltaETfix}.%
Clearly most, but not all, of the differences between the jets found by the
JetClu and Midpoint Algorithms are removed in the fixed version of the latter.
\ The small $R$ ``fix'' suggested for the Midpoint Algorithm can also be
employed for the Seedless Algorithm but, like the Midpoint Algorithm, it will
still miss the middle (now unstable) cone.

Before closing this brief summary of our results, we should say a few more
words about the Run I CDF\ algorithm that we used as a reference. \ In
particular, while ratcheting is difficult to simulate in perturbation theory,
we can attempt to clarify how it fits into the current discussion. \ As noted
above, the JetClu Algorithm is defined so that calorimeter towers initially
found around a seed stay with that cone, even as the center of the cone
migrates due to the iteration of the cone algorithm. \ For the simple scenario
illustrated in Fig.\ \ref{edist2}a we assume that the locations of the partons
are identified as seeds, even when smearing is present. \ To include both
ratcheting and the way it influences the progress of the stable cone search,
we must define 2 scalar functions of the form of Eq. \ref{scalar}, one to
simulate the search for a stable cone starting at $\rho=0$ and the second for
the search starting at $\rho=1.0$. \ The former function is defined to include
the energy within the range $-R\leq\rho\leq+R$ independent of the value of
$r$, while the second function is defined to always include the energy in the
range $1.0-R\leq\rho\leq1.0+R.$ \ Analyzing the two functions defined in this
way suggests, as expected, that the search that begins at the higher energy
seed will always find a stable cone at the location of that seed, independent
of the amount of smearing. \ (If the smearing is small, there is also a stable
cone at the middle location but the search will terminate after finding the
initial, nearby stable cone.) \ The more surprising result arises from
analyzing the second function, which characterizes the search for a stable
cone seeded by the lower energy parton. In the presence of a small amount of
smearing this function indicates stable cones at both the location of the
lower energy parton and at the middle location. \ Thus the corresponding
search finds a stable cone at the position of the seed and again will
terminate before finding the second stable cone. \ When the smearing is large
enough to wash out the stable cone at the second seed, the effect of
ratcheting is to ensure that the search still finds a stable cone at the
middle location suggested by the perturbative result, $r_{3}=z\rho/\left(
1+z\right)  $ (with a precision given by $\sigma\times e^{-\left(
R/\sigma\right)  ^{2}}$). \ This result suggests that the JetClu Algorithm
with ratcheting always identifies either stable cones at the location of the
seeds or finds a stable cone in the middle that can lead to merging (in the
case of large smearing). \ It is presumably just these last configurations
that lead to the remaining difference between the JetClu Algorithm results and
those of the ``fixed'' Midpoint Algorithm illustrated in Fig. \ref{deltaETfix}%
. \ We find that the jets found by the JetClu Algorithm have the largest
$E_{T}$ values of any of the cone jet algorithms, although the JetClu
Algorithm still does not address the full range of splashout corrections.

In conclusion, we have found that the corrections due to the splashout effects
of showering and hadronization result in unexpected differences between cone
jet algorithms applied to perturbative final states and applied to (simulated)
data. With a better understanding of these effects, we have defined steps that
serve to improve the experimental cone algorithms and minimize these
corrections. \ Further studies are required to meet the goal of 1\% agreement
between theoretical and experimental applications of cone algorithms.

\begin{acknowledgments}

This work was supported in part by the DOE under grant DE-FG03-96-ER40956
and the NSF under grant PHY9901946.

\end{acknowledgments}


\begin{thebibliography}{99}
\bibitem{workshops}G. C. Blazey, \textit{et al.}, \textit{Run II Jet Physics:
Proceedings of the Run II QCD and Weak Boson Physics Workshop},
hep-ex/0005012; P. Aurenche, \textit{et al.}, \textit{The QCD and Standard
Model Working Group: Summary Report from Les Houches}, hep-ph/0005114.

\bibitem{longpaper}S.D. Ellis, J. Huston and M.\ T\"{o}nnesmann, \textit{in
preparation}.

\bibitem{Snowmass}J. E. Huth, \textit{et al}., in \textit{Proceedings of
Research Directions for the Decade: Snowmass 1990}, July 1990, edited by E. L.
Berger (World Scientific, Singapore, 1992) p. 134.

\bibitem{ktcluster}S.D. Ellis and D. Soper, Phys. Rev. \textbf{D48}, 3160
(1993); S. Catani, Yu.L. Dokshitzer and B.R. Webber, Phys. Lett.
\textbf{B285}, 291 (1992); S. Catani, Yu.L. Dokshitzer, M.H. Seymour and B.R.
Webber, Nucl. Phys. \textbf{B406}, 187 (1993).

\bibitem{SDERsep}S.D. Ellis, Z. Kunszt and D. Soper, Phys. Rev Lett.
\textbf{69}, 3615 (1992); S.D. Ellis in Proceedings of the \textit{28th
Rencontres de Moriond: QCD\ and High Energy Hadronic Interactions}, March
1993, p. 235; B. Abbott, \textit{et al}., Fermilab Pub-97-242-E (1997).

\bibitem{Seymour}M.H. Seymour, Nucl. Phys. \textbf{B513}, 269 (1998).

\bibitem{Herwig}G. Marchesini, B.R. Webber, G. Abbiendi, I.G. Knowles, M.H.
Seymour, L. Stanco, Computer Phys. Commun. \textbf{67,} 465 (1992) ; G.
Corcella, I.G. Knowles, G. Marchesini, S. Moretti, K. Odagiri, P. Richardson,
M.H. Seymour, B.R. Webber, JHEP \textbf{01, }010 (2001) [hep-ph/9912396].

\bibitem{JetClu}F. Abe et al. (CDF Collaboration), Phys. Rev. \textbf{D45},
1448 (1992).

\bibitem{D0}D\O\ Collaboration, V.M. Abazov, \textit{et al}., hep-ex/0106032.

\bibitem{EKS}S.D. Ellis, Z. Kunszt and D. Soper, Phys. Rev Lett. \textbf{69},
1496 (1992); Phys. Rev Lett. \textbf{64}, 2121 (1990); Phys. Rev.
\textbf{D40}, 2188 (1989); Phys. Rev. Lett. \textbf{62}, 726 (1989).
\end{thebibliography}
\end{document}